\documentclass[12pt,a4paper]{article}

\usepackage{amsbsy,latexsym}

\newcommand{\cd}{\cdot}

\newcommand{\mod}[1]{\vert {#1}\vert}
\newcommand{\pa}{\partial}

\newcommand{\ee}{\mathrm{e}}

\newcommand{\bra}[1]{\langle #1 \vert}
\newcommand{\ket}[1]{\vert #1 \rangle}

\newcommand{\xb}{{\boldsymbol x}}

\begin{document}

\begin{titlepage}
\rightline{PLY-MS-98-11}
\vskip19truemm
\begin{center}{\Large{\textbf{Colour Charges and the Anti-Screening\\
Contribution to the Interquark Potential}}}\\ [8truemm]
\textsc{Martin
Lavelle\footnote{email: m.lavelle@plymouth.ac.uk} and David
McMullan}\footnote{email: d.mcmullan@plymouth.ac.uk}\\ [5truemm]
\textit{School of Mathematics and Statistics\\ The University of
Plymouth\\ Plymouth, PL4 8AA\\ UK} \end{center}

\bigskip\bigskip\bigskip
\begin{quote}
\textbf{Abstract:} Asymptotic freedom arises from the dominance of
anti-screening over screening in non-abelian gauge theories. In this
paper we will present a simple and physically appealing derivation
of the anti-screening contribution to the interquark potential.
Our method allows us to identify the dominant gluonic distribution
around static quarks. Extensions are discussed.
\end{quote}

\end{titlepage}

The discovery of  asymptotic freedom in non-abelian gauge
theories~\cite{Gross:1973th, Politzer:1973um} opened
a route to unifying the quark model of hadron spectroscopy
with the partonic description of high energy scattering.
The QCD beta function
at one loop in pure $SU(N)$ has the form,
$\beta(g)=-{11g^3}/{3(4\pi)^2}$,
the overall sign being responsible for asymptotic freedom.
This result can, in fact, be split up into two parts:
\begin{equation}\label{betasplit}
\beta(g)=-\frac{g^3}{(4\pi)^2}\left[4-\frac13
\right]\,,
\end{equation}
where the larger term (4) is the contribution of anti-screening
effects and the smaller (-1/3) term screens the colour charge.
The dominance of the former effect is responsible for asymptotic
freedom.

The beta function enters all physical quantities and
this split has thus been recognised in many different contexts:
using the background
field method~\cite{Nielsen:1978rm, Hughes:1980ms,
Hughes:1981nw}; through a study of
instantons~\cite{Vainshtein:1982wh};
and in a recent variational
study of QCD~\cite{Brown:1997gm}. The anti-screening
contribution in (\ref{betasplit}) can in all cases be
understood to result
from the contribution of longitudinal gluons, which
correspond to a generalised Coulombic interaction,
while, separately gauge
invariant, transverse gluons generate the
lesser screening term~\cite{Brown:1997nz}.

The one-loop beta function enters the static interquark potential
in the following way~\cite{Fischler:1977yf,
Appelquist:1977tw}
\begin{equation}\label{statpot}
V(r)=-\frac{g^2 C_F}{4\pi r}\left[ 1+
\frac{g^2}{4\pi}\frac{C_A}{2\pi }\left( 4-\frac13
\right) \log(\mu r) \right]\,.
\end{equation}
This result being generally obtained via a one-loop
perturbative calculation (nested inside a Wilson loop). For recent
work on higher loop corrections to the potential we refer to
Ref.~\cite{Peter:1997me}.

\medskip

In this letter we will demonstrate that the potential between
quarks can be directly obtained once the structure of the gluonic
clouds around static quarks is known~\cite{Lavelle:1997ty}. The gluonic
dressing has
a rich structure and this calculation will identify the most significant component.
We thus now briefly consider how glue dresses quarks.

A fundamental characteristic of gauge theories is that not all of
the fields are physical. Gauss' law imposes the requirement of
local gauge invariance on physical states and observables. After
constructing such gauge invariant configurations, their physical
significance needs to be understood. In a non-abelian gauge theory,
colour can only be a well-defined
quantum number on gauge invariant states~\cite{Lavelle:1996tz}. This
simple observation means that we are forced to identify coloured
matter, such as quarks, with gauge invariant products of the form,
$h^{-1}(x)\psi(x)$, where the dressing term must be constructed so as to
transform as $h^{-1}(x)\to h^{-1}(x)U(x)$ under the gauge transformation
where the matter fields transform as $\psi(x)\to U^{-1}(x)\psi(x)$.

In previous papers we have constructed gauge invariant
physical states corresponding to charged
particles~\cite{Lavelle:1997ty}. These have then been submitted to
detailed tests~\cite{Bagan:1997su, Bagan:1998kg} which
supported our identification.
Using these fields we will now see that
the calculation of the dominant contribution to the interquark
potential becomes no more than a tree level calculation. This
opens the way to a detailed analysis of how glue is distributed
around quarks.

For the special case of a
static\footnote{The general, relativistic case may be found
elsewhere~\cite{Bagan:1998kg, india}} quark, the dressing
must~\cite{Bagan:1998kg, india}
also satisfy: $h^{-1}(x)\partial_0 h(x) = gA_0(x)$.
In the Abelian theory these two demands on the
static dressings may be solved
and one finds
\begin{equation}\label{hqed}
  h^{-1}(x)=\exp\left(i\ee\int_{-\infty}^{x^0}\!ds
  \frac{\pa^iF_{i0}}{\nabla^2}(s,\xb)\right)
  \exp\left(-i\ee\frac{\pa_i A_i}{\nabla^2}\right)\,,
\end{equation}
where
\begin{equation}\label{nabladef}
  \frac{\partial_iA_i}{\nabla^2}(x)=
  -\frac1{4\pi}\int\!d^3\boldsymbol{y}
      \frac{\partial_iA_i(x_0,\boldsymbol{y})}{\mod{\boldsymbol{x}-
      \boldsymbol{y}}}\,.
\end{equation}
We note that the temporally non-local term is itself
gauge invariant, while the other term soaks up the gauge
dependence of the matter field. This latter part of the dressing
was initially introduced by Dirac~\cite{Dirac:1955ca}. He
motivated the choice of this particular dressing by
noting that it generates the electric field appropriate to a static
charge. (It is easy to show that this requirement is insensitive to
the presence of the temporally non-local term in Eq.~\ref{hqed}.)

These fields and the generalisation to arbitrary
velocities have been tested in perturbation theory, where it has
been shown that their on-shell Green's functions do not suffer
from infra-red divergences~\cite{Bagan:1998kg}.
We recall~\cite{weinberg:1995}
that QED suffers from two physically distinct
infra-red divergences:
soft and phase divergences. These different aspects of the
infra-red problem are solved by the appropriate generalisations to moving
charges of the two different terms in the
dressing~(\ref{hqed}). In particular the physically important
soft divergences are eliminated by the gauge dependent and
temporally local structure made up of longitudinal photons.

From our above discussion, we would expect that  the QCD
generalisation of this part of the dressing  generates the
dominant, anti-screening part of the potential which is responsible
for asymptotic freedom. At lowest order in the coupling the QCD
generalisation of this dressing structure is the same as in QED
up to the obvious inclusion of
colour\footnote{We use anti-hermitian colour matrices. We thus have:
$\textrm{tr}T^aT^a=-\frac12(N^2-1)=-NC_F$ and
$f_{abc}f_{abd}=N\delta_{cd}=C_A\delta_{cd}$.},
$h^{-1}(x)=1+g\partial_iA^a_i(x)T^a/\nabla^2$, where, as argued above,
we are setting to unity the gauge invariant term. There is a simple
algorithm (see Appendix~A of Ref.~\cite{Lavelle:1997ty}) which
generates the higher order terms in this part of the dressing,
in such a way as to keep the dressed charge gauge invariant.
At order $g^3$ this yields
\begin{equation}\label{gthree}
h^{-1}(x)=  \exp\left( g \chi(x) \right)+O(g^4)\,,
\end{equation}
with $\chi^a T^a=(\chi_1^a+g\chi^a_2+g^2\chi^a_3+\cdots)T^a$ and
\begin{equation}\label{chibits}
  \chi_1^a=\frac{\partial_j A_j^a}{\nabla^2}
  \,,\quad \chi_2^a=f^{abc}\chi^{bc}\quad
\mathrm{and} \quad \chi_3^a=f^{acb}f^{cef} \chi^{efb}
  \,,
\end{equation}
where we have defined
\begin{equation}\label{chitwo}
\chi^{bc}=\frac{\partial_j}{\nabla^2}\left(
\chi_1^bA_j^c+\frac12(\partial_j\chi_1^b)\chi^c_1
\right)\,,
\end{equation}
and
\begin{eqnarray}\label{chithree}
\chi^{efb}&=&\frac{\partial_j}{\nabla^2}
\Big(
\Big.
\chi^{ef}A_j^b
+\frac12 A^e_j\chi_1^f \chi_1^b
-\frac12 \chi^{ef} (\partial_j\chi_1^b )
\nonumber\\
&&\qquad+\frac12 (\partial_j\chi^{ef})\chi_1^b
-\frac16(\partial_j\chi_1^e )\chi_1^f \chi_1^b
\Big.\Big)\,.
\end{eqnarray}
We will see below that this suffices to determine the
anti-screening contribution to the potential at order $g^4$.

To now calculate the potential between such charges,
we take a quark and an antiquark, both
dressed according to Eq.~\ref{gthree}, and study the
expectation value of
the QCD Hamiltonian. The potential is given by the dependence of
the energy on the separation
of the two quarks, $r:=\vert \boldsymbol y - \boldsymbol
y'\vert$.
The potential is thus given by the $r$ dependent part of
\begin{equation}\label{energy}
  V(r)=
  \frac12 \int\!d^3x
  \bra{\bar\psi(y)h(y)h^{-1}(y')\psi(y')}
({E^a_i}^2(x)+{B^a_i}^2(x)) \ket{\bar\psi(y)h(y)h^{-1}(y')\psi(y')}
  \,.
\end{equation}
This reduces to
\begin{equation}\label{potential}
  V(r)=
  -\textrm{tr}\int\!d^3x \bra0 [E^a_i(x),h^{-1}(y)]h(y)
  [E^a_i(x),h^{-1}(y')]h(y') \ket0
  \,,
\end{equation}
where the trace is over colour and we
have used the fact that $B_i^a$ commutes with the
fields in the dressings.

From (\ref{potential}) we obtain at order $g^2$
\begin{equation}\label{gtwo}
  V^{g^2}(r)=
  -g^2\textrm{tr}\int\!d^3x \bra0 [E^a_i(x),
  \chi^d_1(y)]T^dT^b
  [E^a_i(x),\chi^b_1(y')] \ket0
  \,.
\end{equation}
The commutators follow from the fundamental equal time commutator:
$[E_i^a(x),A_j^b(y)]=i\delta^{ab}
\delta(\boldsymbol{x}-\boldsymbol{y})$. From
this we obtain
\begin{equation}\label{gtworesult}
  V^{g^2}(r)=
  -\frac{g^2 N C_F}{4\pi r}
  \,.
\end{equation}
This we recognise as $N$ times the standard result for the
interquark potential~\cite{Fischler:1977yf,  Appelquist:1977tw},
which is a consequence of
the colour singlet \lq mesonic\rq\ states being necessarily
summed over the colours of the dressed quarks. It is also, up to
the colour factor, just the Coulomb potential appropriate to
Electrodynamics.

We can now move on to the $g^4$ term. From (\ref{potential}) we see that,
essentially, we need to calculate the square of $[E^a_i(x),h^{-1}(y)]h(y)$.
After taking the commutators this will yield terms quadratic in
the vector potential. This can then, to this order in $g$, be
evaluated from the free propagator\footnote{Since the fields
are at the same time, there is no subtlety with time ordering}.
The $A_i$ fields can be split up into transverse, $A_i^T$,
and longitudinal components. Only the former are gauge invariant
and therefore only terms quadratic in $A^T$ need to be retained in
the calculation of the potential. Indeed we only keep terms
quadratic in $A$ and, by hand, replace these with transverse fields
since gauge invariance dictates that only this combination can
survive. This is a major simplification
in the calculation: the vast majority of terms are gauge dependent
and may be dropped (with some effort it may be checked that they
cancel).

Expanding $h^{-1}(y)$ we obtain
\begin{equation}\label{expandh}
h^{-1}(y)=1+g\chi_1(y)+g^2(\frac12{\chi_1^2+\chi_2})
+g^3({\frac16\chi_1^3+\frac12\chi_2\chi_1+
\frac12\chi_1\chi_2 +\chi_3})\,.
\end{equation}
The relevant part of the equal time commutators
in (\ref{potential}) is easily seen to be
\begin{equation}\label{need}
[E^a_i(x),h^{-1}(y)]h(y)
=g[E_i^a(x),\chi_1(y)] +g^2[E_i^a(x),\chi_2(y)]
+g^3 [E_i^a(x),\chi_3(y)]
\,,
\end{equation}
since \textit{all} other terms are gauge dependent and so must cancel in the
potential. In fact from (\ref{chitwo}) and (\ref{chithree}) it
is clear that some of the structures in the latter two commutators
are also gauge dependent and may also be dropped. The
terms in (\ref{need}) which are gauge invariant to this order in the coupling
may be easily found to be
\begin{eqnarray}\label{find}
&&\!\!\!\!\!\!\!\!
[E^a_i(x),h^{-1}(y)]h(y)=\frac{i g}{4\pi}T^a
\partial_i^x\frac1{\mod{\boldsymbol{x}-\boldsymbol{y}}}
-\frac{ig^2f_{abc}}{(4\pi)^2}T^c \!
\int\!d^3z\frac1{\mod{\boldsymbol{z}-\boldsymbol{y}}}
\left(\partial_j^z\partial_i^x
\frac1{\mod{\boldsymbol{x}-\boldsymbol{z}}}
\right) {A^b_j}^T\!(z)\nonumber\\
&&\!\!\!+
\frac{ig^3f_{dbc}f_{daf}}{{4\pi}^3}T^c\!
\int\!d^3z\!\!\int\!d^3w
\frac1{\mod{\boldsymbol{z}-\boldsymbol{y}}}
\left(\partial_j^z\frac1{\mod{\boldsymbol{z}-
\boldsymbol{w}}}\right)\!
\left(\partial_k^w
\partial_i^x
\frac1{\mod{\boldsymbol{x}-\boldsymbol{w}}}\right)
{A^f_k}^T\!(w){A^b_j}^T\!(z)
\,.
\end{eqnarray}
We may now substitute this into~(\ref{potential}). To order
$g^4$ there are three terms: the product of the two terms
which are quadratic in the coupling and also the two multiples of
the terms that are linear and cubic in $g$. In both cases the
$x$-integral is trivial, as we may use integration by parts to
combine the $\partial^x_i$'s and obtain terms of the form,
$\nabla^2_x(1/\mod{\boldsymbol x-\boldsymbol z})=-4
\pi\delta(\boldsymbol x-\boldsymbol
z)$. Noting that the
propagator is diagonal in colour, $\langle A^a_iA^b_j\rangle=
\delta^{ab}\langle A_iA_j\rangle$, we find that all of the three
terms in the product are identical and yield
\begin{equation}\label{potint}
  V^{g^4}(r)=
 - \frac{3g^4 C_FC_A}{(4\pi)^3}\int\!d^3z\int\!d^3w
 \frac1{\mod{\boldsymbol z-\boldsymbol w}}
 \left(  \partial_j^z\frac1{\mod{\boldsymbol z-\boldsymbol y}}\right)
 \left(  \partial_k^w\frac1{\mod{\boldsymbol w-\boldsymbol {y}'}}\right)
\langle A_k^T(w)A_j^T(z)\rangle\,.
\end{equation}
We see that, as promised, the calculation of the next to leading
order part of the potential has reduced to an integral over the
tree-level propagator!

The gauge invariant part of the equal time, free propagator
in co-ordinate space is
\begin{equation}\label{prop}
\langle A_k^T(w)A_j^T(z)\rangle=
\frac1{2\pi^2}\frac{(z-w)_j
(z-w)_k}{{\mod{\boldsymbol z-\boldsymbol w}}^4}
\,.
\end{equation}
Thus this part of the potential may be written as
\begin{equation}\label{potintl}
 V^{g^4}(r)=
  \frac{6\pi g^4 C_F C_A}{4} I(r)\,,
\end{equation}
where
\begin{equation}\label{intr}
 I(r)=
  -\frac1{2\pi^2}\int\!\frac{d^3z}{(2\pi)^3}
 \int\!\frac{d^3w}{(2\pi)^3}
 \frac{(w-y)_j(z-y')_k(z-w)j(z-w)_k}{
 \mod{\boldsymbol w-\boldsymbol y}^3
 \mod{\boldsymbol z-\boldsymbol y'}^3
 \mod{\boldsymbol z-\boldsymbol w}^5}
  \,,
\end{equation}
To evaluate this integral it is convenient to perform the
shift, $w\to w+z$, followed by $z\to z+y'$. This ensures that
the $z$-integral is finite. The denominators containing $z$'s
may be combining using Feynman's trick and the integral
can be read off standard tables. This gives the logarithmically
divergent integral
\begin{equation}\label{intrtwo}
 I(r)=
  -\frac1{2\pi^2}\frac1{(4\pi)^{3/2}\Gamma(\frac32)}
 \int\!\frac{d^3w}{(2\pi)^3}
 \left[\frac{2\boldsymbol w\cd \boldsymbol r}{
\mod{\boldsymbol w}^3
 \mod{ \boldsymbol w+\boldsymbol r}^3}
 + \frac{(\boldsymbol w\cd \boldsymbol r)^2}{
\mod{\boldsymbol w}^5
 \mod{ \boldsymbol w+\boldsymbol r}^3}
 \right]
  \,.
\end{equation}
This may be evaluated directly, or, since we
are only interested in the divergence, we can just study its
behaviour in the region of the singularities. E.g., for the first
integral in (\ref{intrtwo}), which has a
singularity at $w\approx -r$, we may shift $w\to w-r$ to find
\begin{equation}\label{exa}
 \int\!\frac{d^3w}{(2\pi)^3}
 \frac{2\boldsymbol w\cd \boldsymbol r}{
\mod{\boldsymbol w}^3
 \mod{ \boldsymbol w+\boldsymbol r}^3}\to
\int\!\frac{d^3w}{(2\pi)^3}
 \frac{2(\boldsymbol w-\boldsymbol r)\cd \boldsymbol r}{
\mod{\boldsymbol w}^3
 \mod{ \boldsymbol w-\boldsymbol r}^3} =-\frac2r
  \int\!\frac{d^3w}{(2\pi)^3}
 \frac1{\mod{\boldsymbol w}^3}
 +\mathrm{finite\ terms}
 \,.
\end{equation}
The divergence of this integral can be thus read off as
$\log(\mu r)/\pi^2 r$, where $\mu$ is a cutoff.
Similarly it may be shown that the second
integral in (\ref{intrtwo}) yields $-4\log(\mu r)/6r\pi^2$.

Putting everything together we obtain the final result for the $g^4$
contribution to the interquark
potential from that part of the dressing which is essential for
gauge invariance
\begin{equation}\label{finalres}
V^{g^4}(r)=-\frac{g^4}{(4\pi)^2}\frac{4NC_FC_A}{2\pi r}\log(\mu
r)\,.
\end{equation}
This we recognise from (\ref{statpot}) as the expected anti-screening contribution
to the potential at next to leading order.

\medskip

The above derivation of the dominant, anti-screening contribution
to the potential is notable for its directness and simplicity. Let
us now briefly discuss its consequences.

We have demonstrated that \textit{the dressing
structure~(\ref{gthree}) is indeed the most physically
significant part of the glue around static quarks}.
We note here the sensitivity~\cite{Haagensen:1997pi} of
this calculation to the form of the glue.
An immediate task is to find the higher order
anti-screening contributions to the potential that would follow from
this minimally gauge invariant part of the dressing:
at $O(g^6)$ this requires a one-loop calculation.
The non-local Coulombic interaction
which this dressing structure generates
and its implications for interquark forces deserve further
study (see also Ref.~\cite{Chen:1998cc}).

Where does the screening
contribution come from? We have neglected the gauge invariant,
temporally non-local part of the dressing (in QED it is
clear that it commutes with the Hamiltonian) and its structure
in QCD and role in the potential needs to be studied.

The dressing used here solely generates a chromo-electric field.
Lattice calculations~\cite{Green:1997be} indicate that, at
least in the
non-perturbative domain, a static quark-antiquark system
also generates a chromo-magnetic field. The form of the gluonic
dressing responsible for such physics needs to be investigated.
The temporally non-local structure may produce this.

We should, however, also note Ref.~\cite{Cahill:1979dq} where it
was argued that a gauge invariant dressing for a quark-antiquark
\lq mesonic\rq\ structure which was local in time yielded the full
potential~(\ref{statpot}). Such structures and their interplay with
the dressings resulting from our programme need further
investigation. In particular it should be studied whether or not
such a dressed \lq meson\rq\ factorises into two of our separately
gauge invariant quarks (at lowest order it does).

This factorisation into constituent charges must break down, as the
potential is only Coulombic at short distances. At larger
separations a linearly rising potential is expected and a mesonic
flux tube structure should emerge. This is fully consistent with
our construction of charges, since
we have shown that there is a fundamental, topological  obstruction
to constructing gauge invariant quarks~\cite{Lavelle:1997ty}.
Such a breakdown of factorisation signals
the boundaries of the applicability of quark models. The
determination of the scale of the breakdown of factorisation
thus presents a central challenge in QCD.

\vfill\eject

\end{document}